\documentclass[pre,aps,amsmath,amssymb,amsfonts,floatfix,superscriptaddress,showpacs,twocolumn,footinbib]{revtex4-1}
\usepackage{graphicx}
\usepackage{amsmath,amsfonts}
\usepackage{xcolor}
\usepackage{color}
\usepackage{mathtools}
\usepackage{eufrak}
\usepackage{pgfplots}
\usepackage{soul}
\usepackage{multirow}

\newcommand{\up}{\uparrow}
\newcommand{\dn}{\downarrow}
\newcommand{\kv}{\ensuremath{\mathbf{k}}}
\newcommand{\qv}{\ensuremath{\mathbf{q}}}

\newcommand{\ch}{\ensuremath{\text{ch}}}
\newcommand{\sz}{\ensuremath{\text{sp}}}

\newcommand{\pp}{\ensuremath{{pp}}}
\newcommand{\ph}{\ensuremath{{ph}}}
\newcommand{\phv}{\ensuremath{\overline{ph}}}

\newcommand{\sing}{\ensuremath{\text{s}}}
\newcommand{\firr}{\ensuremath{\text{firr}}}
\newcommand{\omegap}{\ensuremath{\widetilde{\omega}}}
\newcommand{\red}{\ensuremath{\text{red}}}

\usepackage{tikz}

\usetikzlibrary{calc}
\usetikzlibrary{decorations.pathmorphing}
\usetikzlibrary{decorations.pathreplacing}
\usetikzlibrary{decorations.markings}
\usetikzlibrary{shapes.misc}
\usetikzlibrary{shapes}
\usetikzlibrary{positioning}
\usetikzlibrary{snakes}
\usetikzlibrary{arrows}
\usetikzlibrary{calc,arrows}
\tikzstyle{decision} = [diamond, draw, fill=blue!20, text width=4.5em, text badly centered, node distance=3cm, inner sep=0pt]
\tikzstyle{block} = [rectangle, draw, fill=blue!20, text width=7em, text centered, rounded corners, minimum height=5em]
\tikzstyle{line} = [draw, -latex']
\tikzstyle{cloud} = [draw, ellipse,fill=red!20, node distance=3cm, minimum height=2em]
\tikzstyle{overbrace style}=[decorate,decoration={brace,raise=2mm,amplitude=3pt}]
\tikzstyle{overbrace text style}=[font=\footnotesize, above, pos=.5, yshift=3mm]
\tikzset{snake it/.style={decorate, decoration=snake}}
\usetikzlibrary{3d}
\usepackage{mathtools}
    \tikzset{
            partial ellipse/.style args={#1:#2:#3}{
                        insert path={+ (#1:#3) arc (#1:#2:#3)}
                            }
                        }

\usetikzlibrary{calc}
\tikzset{
            inertial frame/.style = {x={(-20:2cm)}, y={(-160:2cm)}, z={(90:2cm)}},
              local frame/.style = {shift={(local origin)}, x={(40:.7cm)}, y={(150:.7cm)}, z={(105:.7cm)}}
          }

    \tikzset{middlearrow/.style={
                decoration={markings,
                            mark= at position 0.65 with {\arrow{#1}} ,
                                    },
                                            postaction={decorate}
                                                }
                                                }
\tikzset{cross/.style={cross out, draw, 
         minimum size=2*(#1-\pgflinewidth), 
                  inner sep=0pt, outer sep=0pt}}

\def\presuper#1#2%
  {\mathop{}%
   \mathopen{\vphantom{#2}}^{#1}%
   \kern-0.5\scriptspace%
   #2}

\usepackage{hyperref}
\hypersetup{colorlinks=true,breaklinks,linkcolor=blue,urlcolor=blue,citecolor=blue}
\begin{document}

    \pgfmathdeclarefunction{gauss}{2}{%
          \pgfmathparse{1/(#2*sqrt(2*pi))*exp(-((x-#1)^2)/(2*#2^2))}%
          }
    \pgfmathdeclarefunction{mgauss}{2}{%
          \pgfmathparse{-1/(#2*sqrt(2*pi))*exp(-((x-#1)^2)/(2*#2^2))}%
          }
    \pgfmathdeclarefunction{lorentzian}{2}{%
        \pgfmathparse{1/(#2*pi)*((#2)^2)/((x-#1)^2+(#2)^2)}%
          }
    \pgfmathdeclarefunction{mlorentzian}{2}{%
        \pgfmathparse{-1/(#2*pi)*((#2)^2)/((x-#1)^2+(#2)^2)}%
          }

\author{Friedrich Krien}
\affiliation{Jo\v{z}ef Stefan Institute, Jamova 39, SI-1000, Ljubljana, Slovenia}
\author{Angelo Valli}
\affiliation{Institute for Solid State Physics, Vienna University of Technology, 1040 Vienna, Austria}


\title{Parquet-like equations for the Hedin three-leg vertex}

\begin{abstract}
    Taking the competition and the mutual screening of various bosonic fluctuations in correlated electron systems into account
    requires an unbiased approach to the many-body problem.
    One such approach is the self-consistent solution of the parquet equations, whose numerical treatment in lattice systems is however prohibitively expensive.
    In a recent article it was shown that there exists an alternative to the parquet decomposition of the four-point vertex function,
    which classifies the vertex diagrams according to the principle of single-boson exchange (SBE)
    [F. Krien, A. Valli, and M. Capone,~\href{https://link.aps.org/doi/10.1103/PhysRevB.100.155149}{Phys. Rev. B 100, 155149 (2019)}].
    Here we show that the SBE decomposition leads to a closed set of equations for the Hedin three-leg vertex, the polarization, and the electronic self-energy,
    which sums self-consistently the diagrams of the Maki-Thompson type.
    This circumvents the calculation of four-point vertex functions and the inversion of the Bethe-Salpeter equations, which are the two major bottlenecks of the parquet equations.
    The convergence of the calculation scheme starting from a fully irreducible vertex is demonstrated for the Anderson impurity model.
\end{abstract}

\maketitle
\section{Introduction}

Taking nonlocal correlations in electronic systems into account is a challenging task, which is 
necessary to study, for example, unconventional superconductivity~\cite{Orenstein00,Maier05,Otsuki14,Rohringer18,Kitatani19}
or non-Fermi liquid behavior induced by soft collective modes~\cite{Varma02}.
The theoretical tools to investigate these phenomena are however limited,
in particular in the absence of small parameters and in the presence of multiple competing fluctuations.
In some cases, when correlations are sufficiently short-ranged, cluster approaches~\cite{Lichtenstein00,Kotliar01}
or quantum Monte-Carlo techniques~\cite{Scalapino81} can be applied, for recent applications see, for example, Refs.~\cite{Huang18,Brown19,Vucicevic19}.
These methods give direct access to the correlation functions, without having to discern their quantum field theoretical content in terms of Feynman diagrams.
However, they can not capture long-range correlations beyond the finite-size cluster.

In order to reach the thermodynamic limit, it can be more convenient to employ techniques of quantum field theory,  
which allow systematic approximations of one- and two-particle correlation functions.
In fact, the stochastic sampling~\cite{Prokofev07,Vanhoucke10,Kozik10} of a diagrammatic perturbation series even gives access
to numerically exact solutions when the perturbation order is sufficiently bounded~\cite{Iskakov16,Gukelberger17}.
The effects of long-ranged correlations in the two- and three-dimensional Hubbard model
have been studied successfully within the two-particle self-consistent (TPSC) approach~\cite{Vilk97,Bergeron11}
and using diagrammatic extensions of the dynamical mean-field theory (DMFT)~\cite{Georges96}.
Examples are the dynamical vertex approximation (D$\Gamma$A)~\cite{Toschi07} and the dual fermion approach~\cite{Rubtsov08},
see also Ref.~\cite{Rohringer18} for a review. 
These methods brought important insights into, for instance, the absence of a Mott-Hubbard transition~\cite{Schaefer15,vanLoon18-2,Tanaka18}
and high-temperature superconductivity~\cite{Otsuki14,Kitatani19} in the two-dimensional Hubbard model on the square lattice,
the critical properties of the half-filled three-dimensional Hubbard  model~\cite{Rohringer11,Hirschmeier15},
quantum criticality~\cite{Schaefer17,Hirschmeier18}, and Fermi condensation near van Hove singularities~\cite{Yudin14}. 

Despite this success, there remain open questions and problems 
which have not been addressed sufficiently, 
due to intrinsic limitations of the diagrammatic approximations. 
In particular, the D$\Gamma$A and the dual fermion approach have mainly been applied within their respective ladder approximation, 
to include antiferromagnetic correlation effects into the self-energy. 
However, different bosonic fluctuations, and their mutual feedback, 
are not treated on equal footing. 

The underlying challenge here is an unbiased treatment of the vertex corrections,
which in general requires the self-consistent renormalization of the four-point vertex function
and the inversion of the Bethe-Salpeter equations in the presence of the full momentum dependence of the vertex.
The ladder approximation circumvents this by adding nonlocal terms to the electronic self-energy,
but not to the four-point vertex corrections of the underlying DMFT approximation, which leads to only a partial momentum dependence of the full vertex function.
For instance, this discourages the application of the ladder D$\Gamma$A and ladder dual fermion to transport phenomena,
since the optical response is completely unaffected by the vertex corrections 
of the DMFT approximation~\cite{Khurana90}.
In strongly correlated systems the full momentum dependence of the vertex is however crucial for the optical conductivity, even at high temperature~\cite{Brown19,Vucicevic19}.
Due to the missing vertex corrections the ladder approximation is also thermodynamically inconsistent, which leads to multiple values of the total energy~\cite{Krien17}
and inconsistent critical behavior of single- and two-particle quantities~\cite{Janis17}.

Of course, there exists a rigorous formalism for the full renormalization of the vertex function,
which starts from the parquet decomposition of the vertex and requires the self-consistent solution of the parquet equations~\cite{Dominicis64,Dominicis64-2}.
The parquet formalism is in principle unbiased with respect to the dominance of a particular fluctuation mechanism,
in fact, the self-consistent parquet solution is closely related to the functional renormalization group (fRG)~\cite{Kugler18}.
The algorithmic complexity of the parquet equations is however large due to the memory-intensive storage
of various momentum-dependent vertex functions~\cite{Yang09,Tam13,Li16,Eckhardt18,Li19} and the corresponding matrix inversion of the Bethe-Salpeter equations.
It is indeed possible to solve the parquet equations on a small lattice~\cite{Yang09,Tam13,Valli15,Kauch17,Pudleiner19},
which also allows the application of the full D$\Gamma$A scheme to transport phenomena.
This reveals a significant effect of strong antiferromagnetic fluctuations on the optical conductivity~\cite{Kauch19}.
However, due to the severe restrictions on the lattice size it is not possible to study the effect of long-ranged correlations on transport properties.

We propose in this work an alternative to the traditional parquet formalism, which avoids the storage of four-point vertices and the Bethe-Salpeter equations.
The main idea is to obtain a closed set of equations for the Hedin three-leg vertex~\cite{Hedin65}, which depends only on two momentum-energies,
rather than three as the four-point vertex.
The idea of gaining numerical feasibility in this way is therefore reminiscent of the TRILEX approach~\cite{Ayral15,Ayral15-2},
where nonlocal corrections to the Hedin vertex are however neglected.
In our framework the nonlocal corrections are constructed from a set of parquet-like equations for the Hedin vertex.
The latter are based on the notion of irreducibility with respect to the Hubbard interaction,
rather than with respect to pairs of Green's function lines, which underlies the original parquet equations.
It was shown recently in Ref.~\cite{Krien19-2} that this naturally leads to a unique single-boson exchange (SBE) decomposition of the full four-point vertex, whose reducible components are given by the Hedin vertex and the screened interaction, i.e., Maki-Thompson diagrams~\cite{Maki68,Thompson70}.
This representation of the vertex is reminiscent of the partial bosonization techniques used in the context of the fRG,
see, for example, Refs.~\cite{Karrasch08,Husemann09,Denz19} and the references therein.
The SBE decomposition has several promising applications, such as parametrizations of the vertex function,
and in Ref.~\cite{Stepanov19} it was used to obtain an improved TRILEX approximation. 
Moreover, the SBE classification of diagrams completely avoids the vertex divergences of the two-particle self-energy, which arise from the inversion of the Bethe-Salpeter equations~\cite{Schaefer13,Thunstroem18,Chalupa18}. In fact, the building blocks of the parquet-like equations are manifestly non-divergent, as they are connected to physical response functions~\cite{Krien19-2}. 

The purpose of the present work is to show that the SBE decomposition has a corresponding set of equations
which allows the self-consistent reconstruction of the vertex function,
starting from a fully irreducible vertex (analogous to the parquet equations).
We then show that this set of equations can be cast into a three-leg form, which avoids the storage and handling of intermediate four-point vertices.
In accord with the discussion above, the ultimate goal is to study competing fluctuations in lattice models.
Our formalism is a promising candidate for this, because its complexity lies \textit{between} the ladder approximation on the one side,
and the parquet equations on the other.
However, in this work we set a more humble objective of a proof of concept,
by deriving the parquet-like equations for the Hedin vertex and applying them to zero-dimensional impurity models.
This gives a complete overview of the calculation scheme, but requires significantly less numerical resources and implementational precautions~\cite{Chen92,Rohringer12}.


We therefore define the four- and three-point vertices of the Anderson impurity model (AIM) in the following Sec.~\ref{sec:aimvertex}.
In Sec.~\ref{sec:parquet} we derive parquet-like equations for the Hedin vertex starting from the SBE decomposition and formulate the calculation cycle. 
In Sec.~\ref{sec:results} we solve the set of equations in the atomic limit and for the AIM. 
In Sec.~\ref{sec:outlook} we discuss possible approximation schemes for the extension to lattice models. Finally, in Sec.~\ref{sec:conclusions} we draw our conclusions. 

 \begin{figure*}
     \begin{center}
     \includegraphics[width=\textwidth]{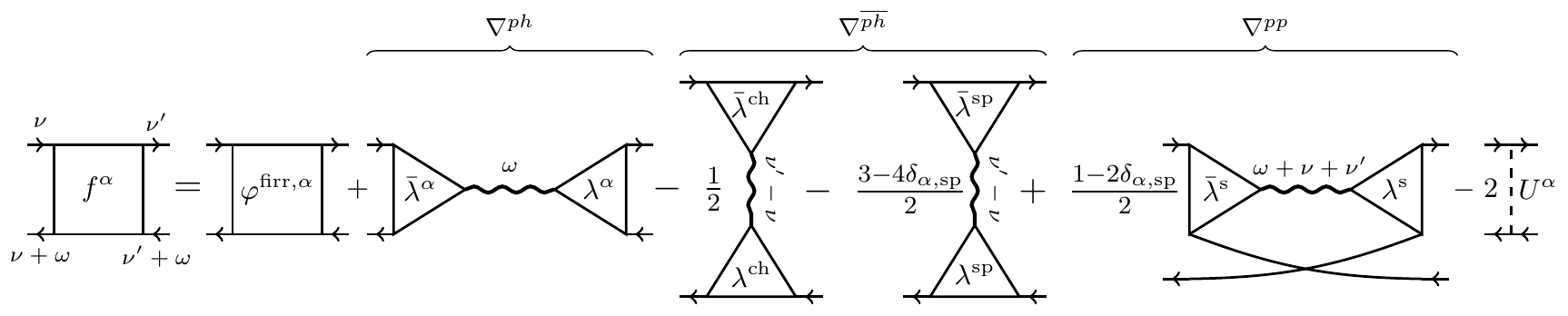}
 \end{center}
     \caption{\label{fig:jib} Feynman graphs corresponding to the SBE decomposition in Eq.~\eqref{eq:jib_full}.
     Triangles denote the Hedin vertices $\lambda$, wiggly lines the screened interaction $w$. See also Ref.~\cite{Krien19-2}.
     The purpose of this work is to show that the fully irreducible vertex $\varphi^\firr$ determines
     all quantities in the figure via a set of self-consistent equations
     (see Figs.~\ref{fig:jibparquet},~\ref{fig:polarization}, and~\ref{fig:dyson}).
     }
     \end{figure*}

\section{Vertex function of the Anderson impurity model}\label{sec:aimvertex}
In our applications we consider the AIM with the imaginary time action,
\begin{align}
  S_{\text{AIM}}=&-\sum_{\nu\sigma}c^*_{\nu\sigma}(\imath\nu+\mu-\Delta_\nu)c^{}_{\nu\sigma}+U \sum_\omega n_{\up\omega} n_{\dn\omega},
  \label{eq:aim}
\end{align}
where $c^*,c$ are Grassmann numbers, $\sigma=\up,\dn$ is the spin index, and $\nu$ and $\omega$ are fermionic and bosonic Matsubara frequencies, respectively.
$U$ is the Hubbard repulsion between the densities $n_{\sigma}=c^*_{\sigma}c^{}_{\sigma}$.
The chemical potential  is fixed to $\mu=\frac{U}{2}$ (half-filling).
Summations over Matsubara frequencies $\nu, \omega$ contain implicitly the factor $T=\beta^{-1}$, the temperature.
We consider two cases for the hybridization function $\Delta_\nu$, which is either set to zero (atomic limit),
or it corresponds to the self-consistent solution of the DMFT equations for the Hubbard model on the square lattice~\footnote{
In this case the hybridization function is fixed via the self-consistency condition, $G_{ii}(\nu)=g(\nu)$,
where $G_{ii}$ is the local lattice Green's function of the Hubbard model in DMFT approximation~\cite{Georges96}.}.
We denote as $g_\sigma(\nu)=-\langle c^{}_{\nu\sigma}c^*_{\nu\sigma}\rangle$ the Green's function of the AIM~\eqref{eq:aim}.
We consider the paramagnetic case, the spin label $\sigma$ is therefore suppressed where unambiguous.

The vertex function of the AIM~\eqref{eq:aim} is the connected part of the four-point correlation function,
\begin{align}
    g^{(4),\alpha}_{\nu\nu'\omega}=&-\frac{1}{2}\sum_{\sigma_i}s^\alpha_{\sigma_1'\sigma_1^{}}s^\alpha_{\sigma_2'\sigma_2^{}}
    \langle{c^{}_{\nu\sigma_1}c^{*}_{\nu+\omega,\sigma_1'}c^{}_{\nu'+\omega,\sigma_2}c^{*}_{\nu'\sigma_2'}}\rangle\notag,
\end{align}
where $s^\alpha$ are the Pauli matrices and the label $\alpha=\ch,\sz$ denotes the charge and spin channel, respectively.
The four-point vertex function $f$ is defined as,
\begin{align}
f^\alpha_{\nu\nu'\omega}=&\frac{g^{(4),\alpha}_{\nu\nu'\omega}-\beta g_\nu g_{\nu+\omega}\delta_{\nu\nu'}
    +2\beta g_\nu g_{\nu'}\delta_{\omega}\delta_{\alpha,\ch}}{g_\nu g_{\nu+\omega}g_{\nu'}g_{\nu'+\omega}}\label{eq:4pvertex}.
\end{align}
The SBE decomposition~\cite{Krien19-2} allows us to express the vertex $f$ in terms of two- and three-point correlation functions,
which are defined as follows. The charge, spin, and singlet susceptibilities are given as,
\begin{align}
    \chi^\alpha_\omega=&-\langle{\rho^\alpha_{-\omega}\rho^\alpha_\omega}\rangle+\beta\langle n\rangle\langle n\rangle\delta_\omega\delta_{\alpha,\ch},\\
    \chi^\sing_{\omegap}=&-\left\langle \rho^-_{-\omegap}\rho^{+}_{\omegap}\right\rangle,
\end{align}
where $\rho^\ch=n_\up+n_\dn=n$ and $\rho^\sz=n_\up-n_\dn$ in the first line are the charge and spin densities 
whereas $\rho^+=c^{*}_\up c^{*}_\dn$ and $\rho^-=c_\dn c_\up$ describe the creation and annihilation of an electron pair~\footnote{
It is in general convenient to differentiate the transferred frequencies $\omega$ and $\omegap$ of particle-hole and particle-particle excitations, respectively.}.
The susceptibility defines the screened interaction,
\begin{eqnarray}
    w^\alpha_\omega=U^\alpha+\frac{1}{2}U^\alpha\chi^\alpha_\omega U^\alpha,\label{eq:w}
\end{eqnarray}
where $U^\ch=U, U^\sz=-U, U^\sing=2U$ is the bare interaction.
The latter corresponds to the leading term of the two-particle self-energy of the respective Bethe-Salpeter equation,
which is uniquely defined, see appendix~\ref{app:fierz}. We define the right-sided Hedin three-leg vertices as,
\begin{align}
    \bar{\lambda}^{\alpha}_{\nu\omega}=\frac{
        \frac{1}{2}\sum_{\sigma\sigma'}s^\alpha_{\sigma'\sigma}\langle{c^{}_{\nu\sigma}c^{*}_{\nu+\omega,\sigma'}\rho^\alpha_\omega}\rangle
        +\beta g_\nu \langle n\rangle\delta_{\omega}\delta_{\alpha,\ch}}
    {g_\nu g_{\nu+\omega}w^\alpha_\omega/U^\alpha},\label{eq:hedinvertex}
\end{align}
for the particle-hole channels, $\alpha=\ch, \sz$, and
\begin{align}
    \bar{\lambda}^{\sing}_{\nu\omegap}=\frac{
    \left\langle c_{\nu\up}c_{\widetilde{\omega}-\nu,\dn}\rho^{+}_{\widetilde{\omega}}\right\rangle
}{g_\nu g_{\omegap-\nu}w^\sing_{\omegap}/U^\sing},\label{eq:lambdasing}
\end{align}
for the singlet particle-particle channel, $\alpha=\sing$.
Due to the factor $w^\alpha/U^\alpha$ in the denominators of Eqs.~\eqref{eq:hedinvertex} and~\eqref{eq:lambdasing}
the Hedin vertices are irreducible with respect to the bare interaction $U^\alpha$ of the corresponding channel, see also Refs.~\cite{Krien19,Krien19-2}.

\section{Parquet-like equations}\label{sec:parquet}
We derive a closed set of equations for the Hedin vertices, the polarization, and the electronic self-energy.
\subsection{SBE decomposition}
The traditional parquet decomposition classifies Feynman diagrams according to their reducible/irreducible property
with respect to pairs of Green's function lines, see, for example, Refs.~\cite{Dominicis64,Dominicis64-2,Bickers04,Rohringer12}.
It was shown in Ref.~\cite{Krien19-2} that, alternatively, diagrams may be grouped according the criterion whether they contain insertions of the bare interaction $U^\alpha$ or not~\footnote{
The notion '$U$-reducible' does \textit{not} imply removal of Green's function legs attached to the bare interaction.
}.
The reference explains the relation between the two notions of reducibility in detail. 
Hence, the SBE decomposition partitions the vertex function into one vertex $\varphi^{\firr}$ that is fully irreducible,
and three vertices $\nabla$ that are reducible, with respect to the bare interaction $U^\alpha$. 
The reducible contributions can be interpreted in terms of the exchange of a single boson. 
The decomposition is exact and given by (see Fig.~\ref{fig:jib}),
\begin{align}
    f^\alpha_{\nu\nu'\omega}\!=\!\varphi^{\firr,\alpha}_{\nu\nu'\omega}\!+\!\nabla^{ph,\alpha}_{\nu\nu'\omega}
    \!+\!\nabla^{\overline{ph},\alpha}_{\nu\nu'\omega}\!+\!\nabla^{\pp,\alpha}_{\nu\nu',\omega+\nu+\nu'}\!-\!2U^\alpha,\label{eq:jib_full}
\end{align}
where $ph$ and $\overline{ph}$ denote the horizontal and vertical particle-hole channels and $pp$ the particle-particle channel,
respectively, $2U^\alpha$ is a double counting correction.
Equation.~\eqref{eq:jib_full} is formulated for the particle-hole notation of the vertex for the SU(2)-symmetric case ($\alpha=\ch,\sz$).

The key feature of the SBE decomposition is that the $U$-reducible vertices $\nabla$ are given by the Hedin three-leg vertex and the screened interaction $w$ (the boson). In the horizontal particle-hole ($U$-$ph$) channel,
\begin{align}
  \nabla^{\ph,\alpha}_{\nu\nu'\omega}=\bar{\lambda}^{\alpha}_{\nu\omega}w^\alpha_{\omega}{\lambda}^{\alpha}_{\nu'\omega},\label{eq:nablaph}
\end{align}
where $\bar{\lambda}$ and ${\lambda}$ are the right- and left-sided Hedin vertices, respectively, which are equal under time-reversal symmetry, $\bar{\lambda}={\lambda}$. The $U$-$\phv$-reducible vertex for the vertical particle-hole channel is obtained by applying the crossing relation,
\begin{align}
    \nabla^{\overline{ph},\alpha}_{\nu\nu'\omega}
    \!=&-\!\frac{1}{2}\!\!\left(\!\nabla^{ph,\ch}_{\nu,\nu+\omega,\nu'-\nu}\!+\![3\!-\!4\delta_{\alpha,\sz}]\nabla^{ph,\sz}_{\nu,\nu+\omega,\nu'-\nu}\!\right)\!,\!\label{eq:nablav}
\end{align}
and finally the $U$-$pp$-reducible vertex is given as,
\begin{align}
    \nabla^{\pp,\alpha}_{\nu\nu'\omegap}=\frac{1-2\delta_{\alpha,\sz}}{2}\bar{\lambda}^{\sing}_{\nu\omegap}w^\sing_{\omegap}\lambda^{\sing}_{\nu'\omegap},                      
    \label{eq:nablapp}
\end{align}
where $\bar{\lambda}^{\sing}={\lambda}^{\sing}$ and $w^\sing$ are a Hedin-like vertex and the screened interaction for the singlet particle-particle channel ($\omegap=\omega+\nu+\nu'$ is the transferred frequency of a particle-particle pair).
\begin{figure*}
    \begin{center}
     \includegraphics[width=\textwidth]{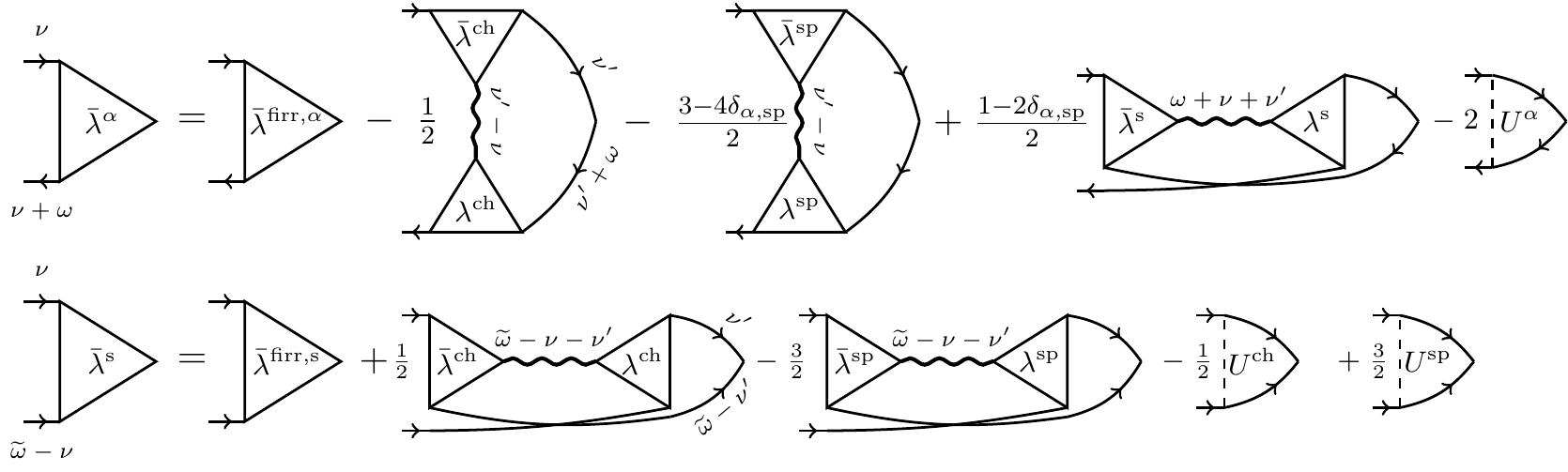}
\end{center}
    \caption{\label{fig:jibparquet}
    Feynman diagrams corresponding to the parquet-like equations for the Hedin vertices $\lambda^{\alpha}$.
    The set of three coupled equations can be solved iteratively for given fully $U$-irreducible three-leg vertices $\lambda^{\firr,\alpha}$.
    Top: Parquet equation for the particle-hole channels, $\bar{\lambda}^{\ch}$ and $\bar{\lambda}^{\sz}$. Bottom: Singlet particle-particle channel, $\bar{\lambda}^{\sing}$.
    }
    \end{figure*}

Next, we write the full vertex $f$ in terms of vertices $\varphi$ which are $U$-irreducible in a particular channel,
\begin{subequations}
\begin{align}
    f^\alpha_{\nu\nu'\omega}=&\varphi^{\ph,\alpha}_{\nu\nu'\omega}+\nabla^{ph,\alpha}_{\nu\nu'\omega},\label{eq:fph}\\
    =&\varphi^{\phv,\alpha}_{\nu\nu'\omega}+\nabla^{\overline{ph},\alpha}_{\nu\nu'\omega},\label{eq:fvph}\\
    =&\varphi^{\pp,\alpha}_{\nu\nu',\omega+\nu+\nu'}+\nabla^{\pp,\alpha}_{\nu\nu',\omega+\nu+\nu'},\label{eq:uspp}
\end{align}
\end{subequations}
Here, $\varphi^{\ph}, \varphi^{\phv}, \varphi^{\pp}$ are $U$-irreducible in the horizontal or vertical particle-hole channel,
or in the particle-particle channel, respectively. 
The SBE decomposition is fully analogous (but not equivalent) to the parquet decomposition~\footnote{ 
The two decompositions are similar because the full vertex is split 
into one contribution which is fully irreducible, and three contributions which are reducible. 
On the other hand, the decompositions are different because their reducible (irreducible) objects have no one-to-one correspondence in terms of Feynman diagrams.}, in fact, Eqs.~\eqref{eq:fph}-\eqref{eq:uspp} play a similar role as the Bethe-Salpeter equations. We combine these equations with Eq.~\eqref{eq:jib_full}, leading to the set of equations,
\begin{subequations}
\begin{align}
    \varphi^{\ph,\alpha}_{\nu\nu'\omega}=\varphi^{\firr,\alpha}_{\nu\nu'\omega}+\nabla^{\phv,\alpha}_{\nu\nu'\omega}
    +\nabla^{\pp,\alpha}_{\nu\nu',\omega+\nu+\nu'}-2U^\alpha,\label{eq:hchannel}\\
    \varphi^{\phv,\alpha}_{\nu\nu'\omega}=\varphi^{\firr,\alpha}_{\nu\nu'\omega}+\nabla^{\ph,\alpha}_{\nu\nu'\omega}
    +\nabla^{\pp,\alpha}_{\nu\nu',\omega+\nu+\nu'}-2U^\alpha,\label{eq:vchannel}\\
    \varphi^{\pp,\alpha}_{\nu\nu',\omega+\nu+\nu'}=\varphi^{\firr,\alpha}_{\nu\nu'\omega}
    +\nabla^{\ph,\alpha}_{\nu\nu'\omega}+\nabla^{\phv,\alpha}_{\nu\nu'\omega}-2U^\alpha\label{eq:schannel}.
\end{align}
\end{subequations}
We stress that Feynman diagrams that contribute to the vertex $\varphi^{r,\alpha}$ do not have bare interaction lines in the channel
$r$, where $r=\ph,\phv,\pp$, but they \textit{can} have such insertions in other channels.
The fully irreducible diagrams represented by $\varphi^{\firr,\alpha}$ do not have bare interaction lines in \textit{any} of the channels~\cite{Krien19-2}.

\subsection{Closed set of equations}
As in the original parquet formalism, the strategy is to use the fully irreducible vertex $\varphi^{\firr}$ as the basis for approximations,
whereas all remaining vertices $\varphi$ and $\nabla$, as well as the electronic self-energy and the polarization function, are determined self-consistently.
To this end, we need to close the set of Eqs.~\eqref{eq:jib_full},~\eqref{eq:hchannel},~\eqref{eq:vchannel},~\eqref{eq:schannel}.

This can indeed be achieved because the vertices $\varphi$ and $\nabla$ are not independent.
They are related via the Hedin vertex, which for $\alpha=\ch, \sz$ is a $U$-$ph$-irreducible three-leg vertex,
\begin{align}
    \bar{\lambda}^{\alpha}_{\nu\omega}=&1+\sum_{\nu'}\varphi^{\ph,\alpha}_{\nu\nu'\omega}g_{\nu'}g_{\nu'+\omega}.\label{eq:lambdafromphiright}
\end{align}

In order to obtain a closed set of equations, it is convenient to eliminate the vertices $\varphi$ which are $U$-irreducible in a particular channel.
To do this for the particle-hole channels, we multiply Eq.~\eqref{eq:hchannel} by $g_{\nu'} g_{\nu'+\omega}$, sum over $\nu'$ and add $1$ on both sides, with the result,
\begin{align}
    \bar{\lambda}^{\alpha}_{\nu\omega}=&\bar{\lambda}^{\firr,\alpha}_{\nu\omega}\label{eq:lambdahchannel}\\
    +&\sum_{\nu'}(\nabla^{\phv,\alpha}_{\nu\nu'\omega}+\nabla^{\pp,\alpha}_{\nu\nu',\omega+\nu+\nu'}-2U^\alpha)g_{\nu'}g_{\nu'+\omega}\notag,
\end{align}
where we used Eq.~\eqref{eq:lambdafromphiright} on the left-hand-side and introduced on the right-hand-side a fully $U$-irreducible three-leg vertex,
\begin{align}
    \bar{\lambda}^{\firr,\alpha}_{\nu\omega}=&1+\sum_{\nu'}\varphi^{\firr,\alpha}_{\nu\nu'\omega}g_{\nu'}g_{\nu'+\omega},\;\;\alpha=\ch, \sz,\label{eq:lambdafirr}
\end{align}
which is not a Hedin vertex, since the vertex part $\varphi^\firr$ is $U$-irreducible with respect to all channels.
Eq.~\eqref{eq:lambdahchannel} is shown in the first line of Fig.~\ref{fig:jibparquet}.
One should note that the Hedin vertex on its left-hand-side also appears on the right-hand-side, but under a summation (or more generally under an integral). 
This integral equation has in fact a simple interpretation.
The Hedin vertex $\bar{\lambda}$ is $U$-$ph$-irreducible, indeed, on the right-hand-side of Eq.~\eqref{eq:lambdahchannel}
the $U$-$ph$-reducible vertex $\nabla^{\ph}$ does not contribute.
However, $\bar{\lambda}$ \textit{does} have $U$-$\phv$- and $U$-$pp$-reducible, as well as fully $U$-irreducible contributions,
the latter are included in $\bar{\lambda}^{\firr,\alpha}$.
These three terms arise on the right-hand-side of Eq.~\eqref{eq:lambdahchannel} 
and the term $-2U^\alpha$ cancels the bare interaction in $\nabla^{\phv}$ and $\nabla^{\pp}$, which would otherwise lead to a $U$-$ph$-reducible contribution.

However, Eq.~\eqref{eq:lambdahchannel} is coupled to the particle-particle channel via $\nabla^{pp}$.
Therefore, we need another equation for the Hedin vertex $\lambda^{\sing}$ of the singlet channel, which is $U$-$pp$-irreducible, 
\begin{align}
    \bar{\lambda}^\sing_{\nu\omegap}=&-1+\frac{1}{2}\sum_{\nu'}\varphi^{\pp,\sing}_{\nu\nu'\omegap}g_{\nu'}g_{\omegap-\nu'}.\label{eq:lambdafromphirightpp}
\end{align}
Similar to the particle-hole case, we can use this relation to eliminate the vertex $\varphi^{\pp}$ in Eq.~\eqref{eq:schannel}.
After some calculation (see Appendix~\ref{app:parquetlike}) we obtain the following relation,
\begin{align}
    \bar{\lambda}^{\sing}_{\nu\omegap}=\bar{\lambda}^{\firr,\sing}_{\nu\omegap}
    +&\frac{1}{2}\sum_{\nu'}\Big[\nabla^{\ph,\ch}_{\nu\nu',\omegap-\nu-\nu'}-3\nabla^{\ph,\sz}_{\nu\nu',\omegap-\nu-\nu'}\notag\\
    -&U^\ch+3U^{\sz}\Big]g_{\nu'}g_{\omegap-\nu'},\label{eq:lambdaschannel}
\end{align}
where we have introduced a fully irreducible vertex $\bar{\lambda}^{\firr,\sing}$ for the singlet channel on the right-hand-side,
\begin{align}
    &\bar{\lambda}^{\firr,\sing}_{\nu\omegap}=\label{eq:lambdafirrsing}\\
    &-1+\frac{1}{4}\sum_{\nu'}\left[\varphi^{\firr,\ch}_{\nu\nu',\omegap-\nu-\nu'}-3\varphi^{\firr,\sz}_{\nu\nu',\omegap-\nu-\nu'}\right]g_{\nu'}g_{\omegap-\nu'}.\notag
\end{align}
Eq.~\eqref{eq:lambdaschannel} is shown in the second line of Fig.~\ref{fig:jibparquet}, its interpretation is similar to Eq.~\eqref{eq:lambdahchannel}:
In the particle-particle channel the Hedin vertex is given by fully $U$-irreducible diagrams, $\bar{\lambda}^{\firr,\sing}$,
and by $U$-$\ph$- and $U$-$\phv$-reducible diagrams, $\nabla^{\ph}$ and $\nabla^{\phv}$.
These vertices contribute equally (see Eq.~\ref{app:parquetsing} in Appendix~\ref{app:parquetlike})
and therefore $\nabla^{\ph}$ is simply counted twice in Eq.~\eqref{eq:lambdaschannel}.

\subsection{Polarization and self-energy}

\begin{figure}
\begin{center}
     \includegraphics[width=0.49\textwidth]{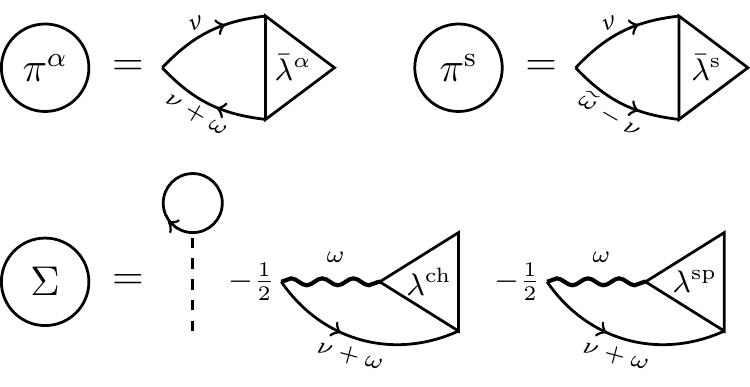}
\end{center}
    \caption{\label{fig:polarization}
    Top: Relation between the polarization and the right-sided Hedin vertex.
    Bottom: Hedin equation.
    }
    \end{figure}

As a last step, we need to introduce a prescription for the renormalization of the fermionic and bosonic lines in Fig.~\ref{fig:jibparquet}.
To this end, we recall that the polarization function $\pi$ is given by the Hedin vertex,
\begin{align}
    \pi^\alpha_\omega=&\sum_\nu g_{\nu}g_{\nu+\omega}\bar{\lambda}^{\alpha}_{\nu\omega},\;\;\alpha=\ch,\sz,\label{eq:pi_ph}\\
    \pi^\sing_{\omegap}=&\sum_\nu g_{\nu}g_{\omegap-\nu}\bar{\lambda}^{\sing}_{\nu\omegap},\label{eq:pi_pp}
\end{align}
which defines the bosonic self-energies. These relations are shown in the first line of Fig.~\ref{fig:polarization}.
The screened interaction is then given as~\footnote{A factor $\frac{1}{2}$ arises in the singlet particle-particle channel due to the indistinguishability of identical particles. For the chosen definitions this factor shows up in equation~\eqref{eq:wpp} for the screened interaction $w^\sing$.},
\begin{subequations}
\begin{align}
  w^\alpha_\omega=&\frac{U^\alpha}{1-U^\alpha\pi^\alpha_\omega},\;\;(\alpha=\ch, \sz),\label{eq:wph}\\
  w^\sing_{\omegap}=&\frac{U^\sing}{1-\frac{1}{2}U^\sing\pi^\sing_{\omegap}}.\label{eq:wpp}
\end{align}
\end{subequations}
These geometric series are formally similar to the random phase approximation (RPA),
but one should keep in mind that here the polarization function $\pi$ is dressed with vertex corrections 
(see also Ref.~\cite{Krien19} and Table I in Ref.~\cite{Krien19-2}). 
Finally, the fermionic self-energy is given by the Hedin vertex through the Hedin equation,
\begin{align}
    \Sigma_\nu=\frac{U\langle n\rangle}{2}-\frac{1}{2}\sum_\omega g_{\nu+\omega}\left[w^\ch_\omega\lambda^\ch_{\nu\omega}+w^\sz_\omega\lambda^\sz_{\nu\omega}\right],
    \label{eq:hedin}
\end{align}
which is shown in the second line of Fig.~\ref{fig:polarization}.

For a model with the Hubbard interaction $Un_\up n_\dn$, the Hedin equation is subjected to the (Fierz) decoupling ambiguity. 
Eq.~\eqref{eq:hedin} corresponds to a symmetric splitting of $U$ into the charge and spin channels,
which is in general the best option~\cite{Zamani16}, since it does not lead to a shift of the chemical potential.
The decoupling ambiguity only affects the single-particle self-energy,
but not the definition of the Hedin vertex or the SBE decomposition, see also Appendix~\ref{app:fierz}.

\begin{figure}\begin{center}
     \includegraphics[width=0.55\textwidth]{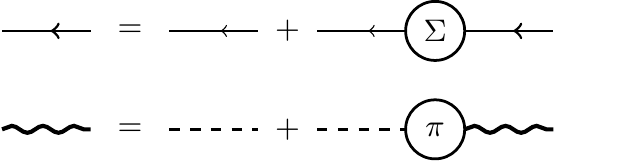}
\end{center}
    \caption{\label{fig:dyson} Dyson equations for the fermionic (top) and bosonic (bottom) propagators.
    Thin arrows and dashed lines represent the bare propagators $g^0$ and $U^\alpha$, respectively.
    }
    \end{figure}

The Hedin equation~\eqref{eq:hedin} determines the Green's function via the Dyson equation,
\begin{align}
    g_\nu=\frac{g_\nu^0}{1-g_\nu^0\Sigma_\nu}\label{eq:dyson}
\end{align}
where $g^0=[\imath\nu-\Delta_\nu+\mu]^{-1}$ is the non-interacting Green's function.
The hybridization function $\Delta$ and the chemical potential $\mu$ allow to control the non-interacting band and the filling, respectively. 
Fig.~\ref{fig:dyson} shows diagrammatic representations of the Dyson equations~\eqref{eq:wph},~\eqref{eq:wpp}, and~\eqref{eq:dyson}.

\subsection{Calculation cycle}\label{sec:cycle}
\begin{figure}[b!]
    \begin{center}
        \includegraphics[width=0.5\textwidth]{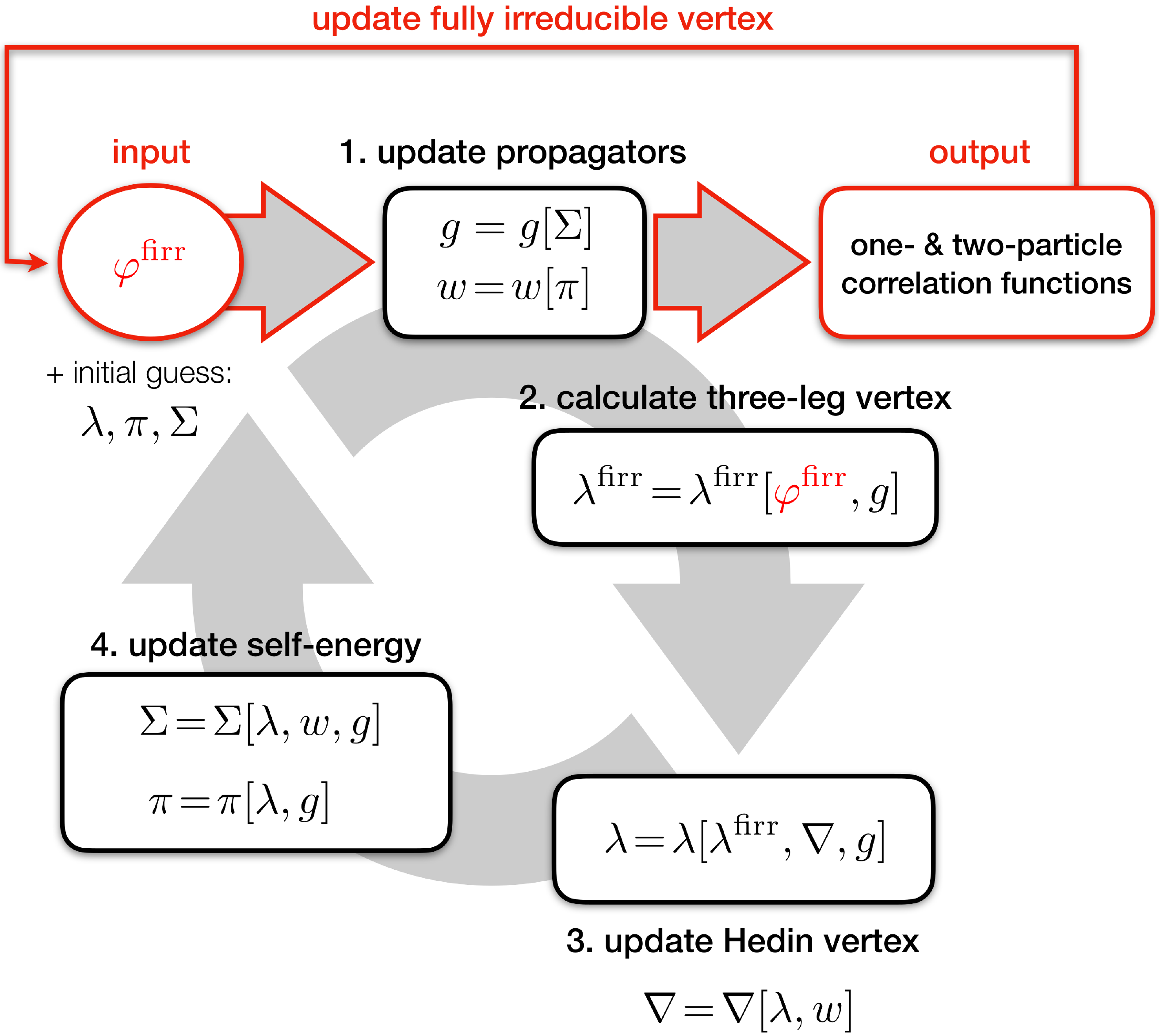};
    \end{center}
    \vspace{-0.5cm}
    \caption{\label{fig:cycle} (Color online) Three-leg parquet self-consistent cycle. 
    Given a fully $U$-irreducible vertex $\varphi^{\firr}$, the parquet equations are iterated self-consistently following four steps (see Sec.~\ref{sec:cycle}) to yield all one- and two-particle correlation functions. Highlighted (in red) are the parquet inputs and outputs 
    as well as the external self-consistency cycle to update $\varphi^{\firr}$, which is discussed in Sec.~\ref{sec:sbedga}. 
    }
\end{figure}

For a given fully $U$-irreducible vertex $\varphi^{\firr}$, the parquet-like equations in Fig.~\ref{fig:jibparquet},
the update formulas for the self-energies in Fig.~\ref{fig:polarization},
and the Dyson equations in Fig.~\ref{fig:dyson} form a closed set of equations.
We formulate the calculation cycle for the SBE equations, which is shown in Fig.~\ref{fig:cycle}.

\noindent \textit{Step 0: Approximation and initial guess.} 
The parquet cycle requires an approximation for the fully $U$-irreducible vertex $\varphi^\firr$,
which remains fixed through the iterations and is not renormalized. 
Furthermore, an initial guess for the Hedin vertices $\lambda$ and for the self-energies $\pi$ and $\Sigma$ is required, which are updated at each iteration.
For example, to begin the calculation from the non-interacting limit, one sets $\lambda^\ch=\lambda^\sz=1, \lambda^\sing=-1, \Sigma=0$,
and the polarizations are set to $\pi^\ch(\omega)=\pi^\sz(\omega)=\sum_\nu g^0_\nu g^0_{\nu+\omega}$ and $\pi^\sing(\omegap)=-\sum_{\nu}g^0_\nu g^0_{\omegap-\nu}$.

\noindent \textit{Step 1: Update propagators.} 
The bosonic and fermionic propagators $w$ and $g$ are updated using the Dyson equations~\eqref{eq:wph},~\eqref{eq:wpp}, and~\eqref{eq:dyson}. 

\noindent \textit{Step 2: Calculate fully irreducible three-leg vertex.} 
The fully irreducible vertex $\varphi^\firr$ is converted into the three-leg vertices $\lambda^\firr$ using Eqs.~\eqref{eq:lambdafirr} and~\eqref{eq:lambdafirrsing}~\footnote{
Step 2 of the calculation cycle is necessary in order to update the Green's function legs in~\eqref{eq:lambdafirr} and~\eqref{eq:lambdafirrsing}.
One may apply approximations directly to the three-leg objects $\lambda^\firr$. In this case, Step 2 can be omitted.}.

\noindent \textit{Step 3: Update Hedin vertices.} The Hedin vertices $\lambda$ are updated using the parquet-like formulas~\eqref{eq:lambdahchannel} and~\eqref{eq:lambdaschannel}.

\noindent \textit{Step 4: Update self-energies.} The polarization $\pi$ and the fermionic self-energy $\Sigma$ are calculated from Eqs.~\eqref{eq:pi_ph}, \eqref{eq:pi_pp}, and \eqref{eq:hedin}, respectively. 

\noindent Steps from 1 to 4 are iterated until convergence. 

A simple iteration of the steps above, even with a linear mixing of the quantities $\Sigma$, $\pi$, and $\lambda$ with their values from previous iterations, is not in general the most efficient solution scheme. 
In order to improve the convergence, a non-linear Broyden root-finding algorithm can be used. 
For the same number of linear updates, the Broyden solver typically converges faster than a simple  mixing scheme and enhances the stability of the algorithm, see, e.g., Ref.~~\cite{Zitko09}.

\section{Numerical examples}\label{sec:results}
\subsection{Atomic limit}\label{sec:results_AL}
The action of the atomic limit (AL) is obtained from Eq.~\eqref{eq:aim} 
by setting $\Delta_{\nu}=0$. 
The main advantage of considering the AL is that one can derive exact 
analytical expressions for the vertex functions~\cite{Thunstroem18}. 
In particular, with the knowledge of the analytical form of the full vertex $f$, 
which can be found in Ref.~\cite{Thunstroem18}, and of the Hedin vertex $\lambda$ (see Appendix~\ref{app:hedin}), 
the SBE decomposition~\eqref{eq:jib_full} yields the exact fully $U$-irreducible vertex $\varphi^{\firr}$ of the AL. 
This allows us to test the parquet equations for the Hedin vertex in a controlled environment by feeding the $\varphi^{\firr}$ of the AL as an input, and benchmark the resulting one- and two-particle correlation functions against the exact solution. 
Furthermore, since there is no other energy scale beside the bare interaction $U$ and the temperature $T$, the AL can be characterized in terms of the ratio $U/T$ alone. 

\begin{figure}
    \begin{center}
        \includegraphics[width=0.47\textwidth]{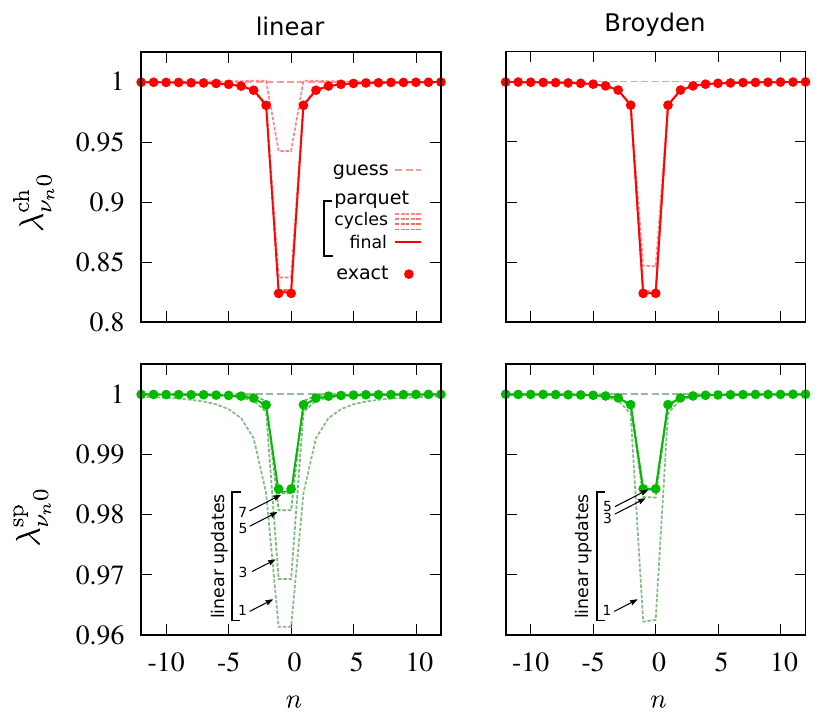}
    \end{center}
    \vspace{-0.5cm}
    \caption{\label{fig:converge_lnu0} (Color online) Convergence of the self-consistent parquet cycle 
    for the AL Hedin vertex $\lambda^\alpha_{\nu_n \omega_m}$ ($\alpha = \ch, \sz$) for $\omega_m=0$ at $U/T=2$. 
    From the initial guess $\lambda^{\ch,\sz}=1$ (long-dashed line), 
    the vertex flows to the exact solution (symbols) in a few iterations 
    with a linear mixing scheme (left panel). The convergence accuracy and speed can be improved using a non-linear Broyden update at each iteration (right panel). 
    The particle-particle vertex in the singlet channel is obtained by symmetry $\lambda^{\sing}_{\nu\omega}=-\lambda^{\ch}_{\nu,-\omega}$ (not shown). 
    }
\end{figure}

As a proof of concept, we begin by considering the weak-coupling regime, at $U/T=2$. 
Fig.~\ref{fig:converge_lnu0} shows the convergence of the Hedin vertex $\lambda^{\alpha}_{\nu\omega=0}$, $\alpha=\ch, \sz, \sing$. 
In this regime, we can take an agnostic guess and set 
$\lambda^{\ch}=\lambda^{\sz}=1$, $\lambda^{\sing}=-1$, and the self-energies of the non-interacting limit, as discussed in Sec.~\ref{sec:cycle}. We can observe the Hedin vertex  flowing from $\pm 1$ towards the exact solution within a few iteration already with a simple linear update scheme (left panels of Fig.~\ref{fig:converge_lnu0}). The accuracy and the speed of the convergence can further be improved by using a non-linear Broyden update at each iteration. The most significant systematic error of the calculation is determined by the size of the frequency window chosen for the calculation. However, the relatively high-accuracy results shown in  Fig.~\ref{fig:converge_lnu0} can already be obtained with an unimpressive  $(N_{\nu},N_{\omega})=(32,16)$ number of Matsubara frequencies, and the calculation can be converged on a \textit{single core} within a few minutes. 
Calculations at larger $U/T$, as those shown in the following, were performed with a wider frequency windows, up to $(N_{\nu},N_{\omega})=(128,32)$, to achieve a comparable accuracy. 
It is noteworthy that the computational cost for the solution of the three-leg parquet equations  scales $\propto (N_{\nu} N_k)^2(N_{\omega} N_q)$ for linear updates, 
where the parameters $N_k$ and $N_q$ determine the mesh for the lattice momenta (with $N_k=N_q=1$ for the AL and the AIM, see also Sec.~\ref{sec:outlook}). 
In contrast, since the solution of the standard parquet equations requires the inversion of the Bethe-Salpeter equations, it scales $\propto (N_{\nu}N_{k})^3 N_{\omega}N_q$~\cite{Yang09}. 


\begin{figure}[h!]
    \begin{center}
        \includegraphics[width=0.37\textwidth]{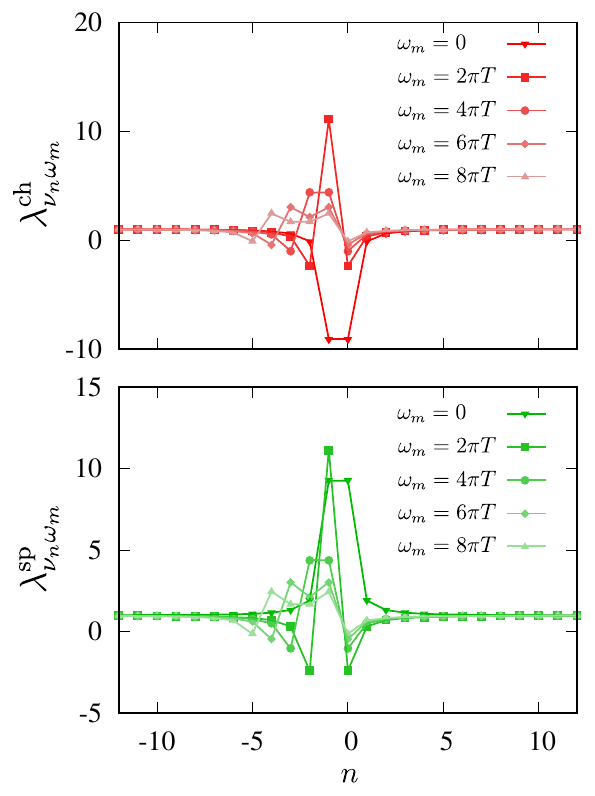}
    \end{center}
    \vspace{-0.5cm}
    \caption{\label{fig:hedin_AL} (Color online) Exact (symbols) 
    and converged  parquet (lines) 
    Hedin vertex $\lambda^{\alpha}_{\nu_n \omega_m}$ ($\alpha = \ch, \sz$) of the AL at $U/T=20$. 
    The particle-particle vertex in the singlet channel is obtained by symmetry $\lambda^{\sing}_{\nu\omega}=-\lambda^{\ch}_{\nu,-\omega}$ (not shown). 
    }
\end{figure}

\begin{figure}[h!]
    \begin{center}
        \includegraphics[width=0.37\textwidth]{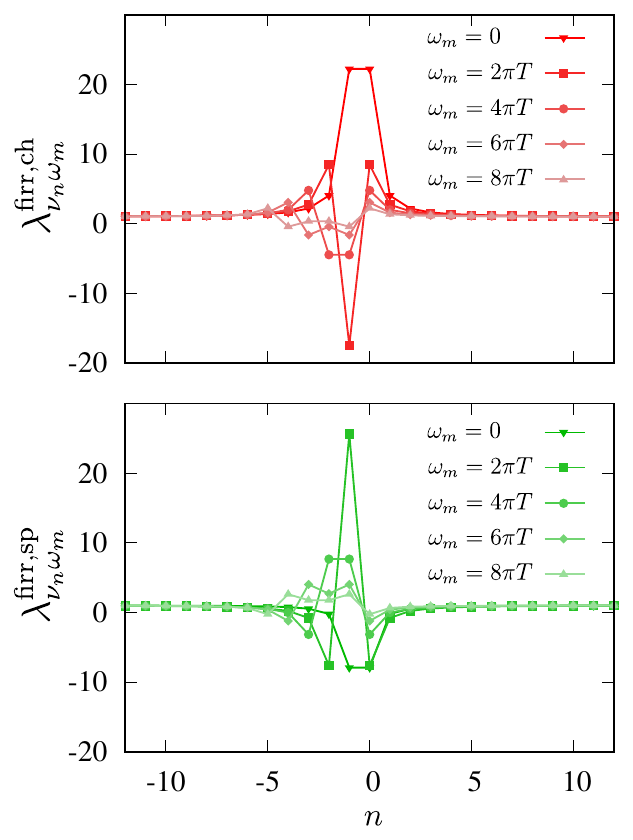}
    \end{center}
    \vspace{-0.5cm}
    \caption{\label{fig:hedin_firr_AL} (Color online) Exact fully irreducible three-leg vertex $\lambda^{\firr,\alpha}_{\nu_n \omega_m}$ ($\alpha = \ch, \sz$) of the AL at $U/T=20$ 
    corresponding to the Hedin vertex in Fig.~\ref{fig:hedin_AL}. 
    }
\end{figure}

In Fig.~\ref{fig:hedin_AL} we compare the exact and the converged parquet Hedin vertex of the AL at $U/T=20$. The structure of the vertex is completely reproduced by the parquet. 
In Fig.~\ref{fig:hedin_firr_AL} we also show the corresponding exact fully irreducible three-leg vertex.
In general, the convergence gets harder upon increasing $U/T$ due to the exponential difference in the fluctuations in the different channels 
$\chi^{\ch}_{\omega=0}/\chi^{\sz}_{\omega=0} \sim e^{-\frac{1}{2}\beta U}$. 
Technical improvements, such as a non-linear Broyden update greatly help the convergence, as already mentioned (see Fig.~\ref{fig:converge_lnu0}). Besides that, it is also important to provide a reasonable initial guess, as for $U/T \gtrsim 2$ the non-interacting limit is quite inconvenient in the case of the AL. In this respect, the knowledge of the exact $\lambda$ of the AL allow us to perform an \textit{annealing} procedure, by feeding as an initial guess of the parquet the exact Hedin vertex and the self-energies for similar parameters. 
On the other hand, the systematic error due to the finite frequency cutoff is rapidly suppressed by increasing the frequency window. 
Also, the tail handling of the Hedin vertex and of the screened interaction is trivial,
since their asymptotic behavior is known exactly~\cite{Wentzell16}, $\lambda^\alpha\rightarrow\pm 1$ and $w^\alpha\rightarrow U^\alpha$.
This is a substantial simplification compared to the traditional parquet equations~\cite{Li16}.
In general, the AIM is expected to display a smoother behavior of the physical quantities, and our numerical calculations presented in Sec.~\ref{sec:results_aim} confirm this expectation. 

\subsection{Anderson impurity model}\label{sec:results_aim}
We consider an AIM corresponding to the self-consistent DMFT solution of the two-dimensional Hubbard model on a square lattice with nearest neighbor hopping $t=1$, 
at half-filling. The correlation functions defined in Sec.~\ref{sec:aimvertex} that correspond to the particle-hole channels ($\alpha=\ch,\sz$)
were evaluated using the ALPS solver~\cite{ALPS2} with improved estimators~\cite{Hafermann12}. 
At half-filling, the Hedin vertex in the singlet particle-particle channel 
can be obtained by symmetry as $\lambda^{\sing}_{\nu\omega}=-\lambda^{\ch}_{\nu,-\omega}$. 
This property is consistent with the numerical results obtained by evaluating the screened interaction $w^\sing$ and $\lambda^\sing$ obtained using the worm sampling method of the w2dynamics package~\cite{Gunacker15,Gunacker16,Wallerberger19}. 

\begin{figure}
    \begin{center}
        \includegraphics[width=0.37\textwidth]{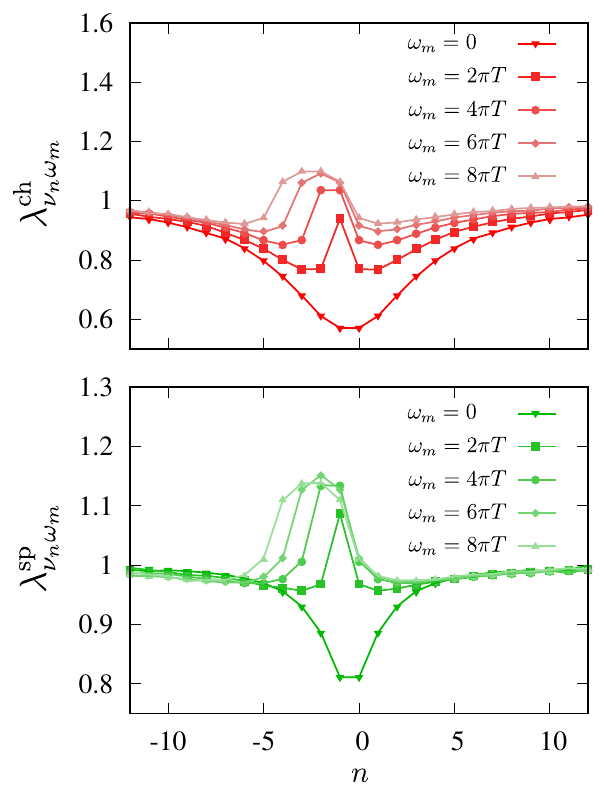}
    \end{center}
    \vspace{-0.5cm}
    \caption{\label{fig:hedin_AIM} (Color online) Exact (symbols) and reconstructed parquet (lines) 
    Hedin vertex $\lambda^{\alpha}_{\nu_n \omega_m}$ ($\alpha = \ch, \sz$) of the AIM at $U=4$ and $\beta=5$ 
    (i.e., $U/T=20$) with $\eta=0.25$ (see text).
    The particle-particle vertex in the singlet channel is obtained by symmetry $\lambda^{\sing}_{\nu\omega}=-\lambda^{\ch}_{\nu,-\omega}$ (not shown). 
    }
\end{figure}

In Fig.~\ref{fig:hedin_AIM} we show the exact Hedin vertex of the AIM at $U=4$ and $\beta=5$. With this choice of parameters, the corresponding lattice Hubbard model lies slightly below the N\'eel temperature of DMFT. It is also interesting to compare the Hedin vertex of the AIM to the one of the AL at the same ratio $U/T=20$ (shown in Fig.~\ref{fig:hedin_AL}). The presence of the self-consistent DMFT bath, describing a Hubbard model in the Fermi liquid regime, strongly suppresses and smoothens the features of the vertex with respect to the AL.

\begin{figure}
    \begin{center}
        \includegraphics[width=0.37\textwidth]{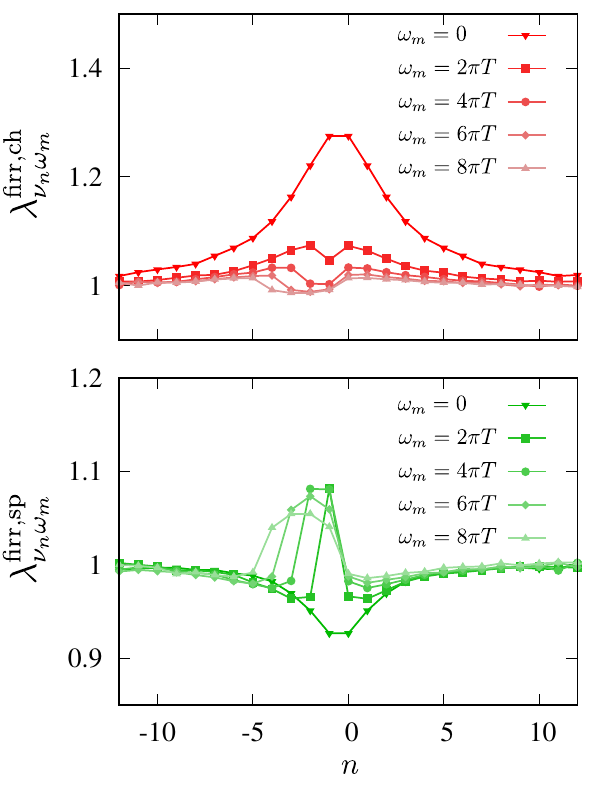}
    \end{center}
    \vspace{-0.5cm}
    \caption{\label{fig:hedin_firr_AIM} (Color online) Exact fully irreducible three-leg vertex $\lambda^{\firr,\alpha}_{\nu_n \omega_m}$ ($\alpha = \ch, \sz$) of the AIM at $U=4$ and $\beta=5$, 
    corresponding to the Hedin vertex in Fig.~\ref{fig:hedin_AIM}. 
    }
\end{figure}

We also numerically demonstrate that the exact Hedin vertex $\lambda$ is an attractive fixed point of the parquet equations. This is done as follows. We provide as input to the parquet the exact 
$\varphi^{\firr}$ obtained from the impurity solver and Eq.~\eqref{eq:jib_full} 
as well as the exact self-energies $\Sigma$ and $\pi$, which we do \textit{not} update in the cycle. 
Furthermore, we need to provide an initial guess for $\lambda$. 
Given $\varphi^{\firr}$, Eq.~\eqref{eq:lambdafirr} yields the fully irreducible three-leg vertex $\lambda^{\firr,\alpha}$ in the particle-hole channels $\alpha = \ch, \sz$. Note that since the self-energy (and hence the Green's function) will not be updated, $\lambda^{\firr,\alpha}$ will also remain identical through the iterations, in contrast to the full parquet solution. The fully irreducible Hedin vertex is shown in Fig.~\ref{fig:hedin_firr_AIM}. 
Then, we iterate Eqs.~\eqref{eq:lambdahchannel} and \eqref{eq:lambdaschannel} and let the Hedin vertex flow towards the fixed point. 

In order to test the flow, we choose as initial guess for $\lambda$ the DMFT (i.e., the numerically exact) vertex of the AIM with a perturbation. 
For the results shown in Fig.~\ref{fig:hedin_AIM}, the perturbation is a frequency-dependent random noise $\eta_{\nu\omega}$ uniformly distributed in a range $[-\eta,\eta]$, so that the input vertex reads 
\begin{equation}
 \lambda^{\alpha}_{\nu\omega} = \lambda^{\alpha,\text{DMFT}}_{\nu\omega} (1-\eta_{\nu\omega}).
\end{equation}
The data in Fig.~\ref{fig:hedin_AIM} are obtained with $\eta=0.25$, iterating the equations as discussed above. The resulting Hedin vertex completely reproduces the exact result, thus demonstrating the stability of the attractive fixed point. 
We verified that the flow is stable for several values of $\eta$ and, quite impressively, even at extreme signal-to-noise ratios such as $\eta=0.99$. 
On the other hand, for the parameters considered, the non-interacting limit is a poor initial guess and the parquet does not converge from there to the numerically exact result. 

\begin{figure}
    \begin{center}
        \includegraphics[width=0.47\textwidth]{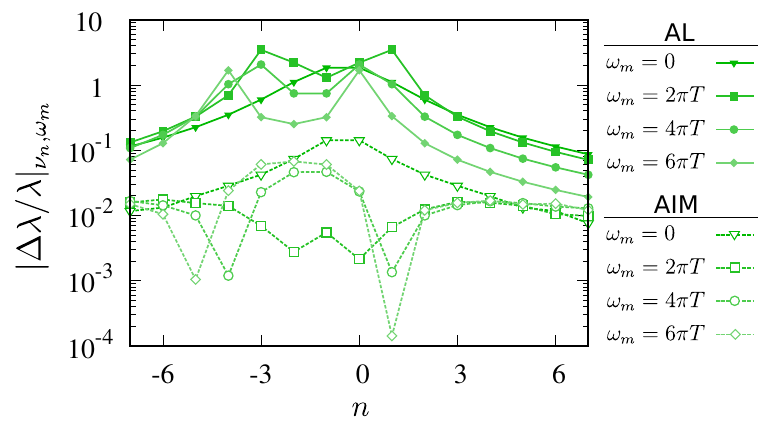}
    \end{center}
    \vspace{-0.5cm}
    \caption{\label{fig:hedin_nabla} (Color online) Relative weight of the $U$-reducible contributions to the spin Hedin vertex, defined as $|\Delta\lambda/\lambda|_{\nu_n,\omega_m} \!=\! |\lambda^{\sz}-\lambda^{\firr,\sz}|/|\lambda^{\sz}|_{\nu_n,\omega_m}$ 
    for the AL (solid line, filled symbols) and the AIM (dashed line, open symbols) at $U=4$ and $\beta=5$ (i.e., $U/T=20$). 
    }
\end{figure}

Finally, by reviewing all our numerical results for the AL (Figs.~\ref{fig:hedin_AL} and \ref{fig:hedin_firr_AL})  and the AIM (Figs.~\ref{fig:hedin_AIM} and \ref{fig:hedin_firr_AIM}), we can comment on the contribution of the $U$-reducible diagrams to the Hedin vertex. This contribution is in general large near an instability. This follows from the definition of the screened interaction in Eq.~\eqref{eq:w}, which implies that for a large impurity susceptibility the corresponding $U$-reducible diagrams in Eq.~\eqref{eq:nablaph} are also large. Of the considered systems, the AL is closer to an instability than the AIM, as $\chi^{\sz}_{\text{AL}}\rightarrow \infty$, whereas $\chi^{\sz}_{\text{AIM}}$ remains finite at $T = 0$ as the local moment is screened by the finite hybridization function. Hence, if $U$-reducible diagrams are important near an instability, there should be a crucial difference in their contribution to the spin Hedin vertex between the AL and the AIM. That this indeed the case is demonstrated in Fig.~\ref{fig:hedin_nabla}, where we compare the relative weight of the $U$-reducible diagrams to the full Hedin vertex $\lambda$.

\section{Outlook: generalization to lattice model}\label{sec:outlook}
So far we have validated the parquet equations for the Hedin vertex for the cases of the AL and the AIM, where the \textit{exact} $\varphi^{\firr}$ can be calculated analytically and numerically, respectively. 
However, in order to address lattice models, for which this information is not available, one needs to rely on suitable \textit{approximations} for $\varphi^{\firr}$ to feed to the parquet solver, 
as we will discuss in the following. 
 
To this end, we generalize the SBE equations to the lattice Hubbard model with the following prescriptions. Any four-point vertex is a function of three momentum-energies, $v(\nu,\nu',\omega)\rightarrow V(k,k',q)$,
where $k=(\kv,\nu)$, $q=(\qv,\omega)$ and $\kv,\qv$ are fermionic and bosonic lattice momentum, respectively.
Any three-leg vertex is a function of two momentum-energies, $\lambda(\nu,\omega)\rightarrow\Lambda(k,q)$,
and for the fermionic and bosonic self-energies, $\Sigma(\nu)\rightarrow\Sigma(k), \pi(\omega)\rightarrow\Pi(q)$,
and likewise for the propagators, $g(\nu)\rightarrow G(k), w(\omega)\rightarrow W(q)$.

\subsection{SBE approximation}\label{sec:SBEapprox}
The lowest-order approximation for the fully irreducible vertex consists in neglecting its frequency dependence, while keeping only the static contributions. 
Within the SBE formalism, this corresponds to 
\begin{align}
    \varphi^{\firr}_{\text{lattice}}(k,k',q)\equiv 0,\label{eq:sbeapprox}
\end{align}
since the bare interaction is already included in the reducible vertex $\nabla$~\cite{Krien19-2}.
This is reminiscent of the \textit{parquet approximation} in the standard parquet formalism. 
The SBE approximation can be considered the lowest-order approximation for the fully irreducible vertex and it is expected to yield a reasonable description of the lattice Hubbard model in the weak-coupling regime. 
The main advantage of this approach is that it allows for a self-consistent solution of the parquet equations on the lattice without any external input, and it does \textit{not} require the calculation of any four-point vertex.
It is plausible that the SBE approximation can not be converged in all parameter regimes, similar to the parquet approximation.
Indeed, in first tests of this approximation in the AL we found that it converges only for small enough values of the ratio $U/T$ (results not shown).

\subsection{Dynamical vertex approximation}\label{sec:sbedga}

Any step beyond the SBE approximation for the fully irreducible vertex requires us to take into account the frequency structure of $\varphi^{\firr}$. 
A possible approximation scheme is inspired by the original D$\Gamma$A, where the fully irreducible vertex is approximated by its \textit{local} contributions, 
obtained from an exactly solvable auxiliary AIM, 
\begin{align}
    \varphi^{\firr}_{\text{lattice}}(k,k',q)\equiv\varphi^{\firr}_{\text{AIM}}(\nu,\nu',\omega).\label{eq:sbedga}
\end{align}
The hybridization function of the AIM can correspond to the DMFT solution of the corresponding lattice problem, but it may also be updated with a suitable outer self-consistency of the calculation cycle in Sec.~\ref{sec:cycle}. Once the parquet cycle is completed, 
it determines all one- and two-particle correlation functions of the lattice Hubbard model generated by the fully irreducible local vertex. 
The SBE-D$\Gamma$A~\eqref{eq:sbedga} is inspired by the original D$\Gamma$A introduced in Ref.~\cite{Toschi07} but its diagrammatic content, implementation details, and algorithmic complexity are different. 
In particular, the SBE-D$\Gamma$A recovers a smaller number of nonlocal diagrams than the D$\Gamma$A,
namely those which correspond to single-boson exchange (cf. Maki-Thompson diagrams~\cite{Maki68,Thompson70}),
while the D$\Gamma$A also accounts for multi-boson exchange (cf. Aslamazov-Larkin diagrams~\cite{Aslamazov68}).
On the other hand, the SBE-D$\Gamma$A is unaffected by divergences of the two-particle self-energy~\cite{Schaefer13,Chalupa18} and has a significantly improved numerical feasibility.
Similar to approximations based on the traditional parquet formalism,
the SBE-D$\Gamma$A satisfies the crossing-symmetry of the vertex function and several more exact relations, see also Appendixes~\ref{app:symmetry} and~\ref{app:exactrelations}. The crossing-symmetry implies the thermodynamic consistency of the potential energy~\footnote{
    The thermodynamic consistency of the \textit{potential} energy in parquet approaches contrasts with the consistency of the \textit{kinetic} energy
    in conserving theories~\cite{Baym62,Krien17,Krienthesis}.} and the Pauli principle~\cite{Bickers04}.
In some approximations the Pauli principle implies at least the partial satisfaction of the Mermin-Wagner theorem~\cite{Vilk97,Kontani06,Katanin09,Rohringer16}.

The underlying assumption of the locality of $\varphi^{\firr}$ raises the question whether the nonlocal correlations of the Hubbard model are indeed captured by the $U$-reducible diagrams generated by the parquet. It is difficult to address this question in general. However, as we explained above in Sec.~\ref{sec:results}, the $U$-reducible diagrams indeed play a key role near some instabilities. This implies that, in order to capture second order phase transitions of the Hubbard model with a nonlocal order parameter, it is crucial to take the spatial dependence of the $U$-reducible diagrams into account~\cite{Rohringer18}.
The SBE-D$\Gamma$A is designed to do precisely this. 
Nevertheless, fully irreducible diagrams may also contribute to nonlocal correlations.
In this case, one could supplement the local fully irreducible vertex in Eq.~\eqref{eq:sbedga}
self-consistently with nonlocal multi-boson contributions,
which are not generated by the three-leg parquet but can be crucial away from particle-hole symmetry~\cite{Bergeron11,Kitatani19}.
We note that, in order to optimize the decay of Matsubara summations,
it is also appealing to apply our formalism to dual fermions~\cite{Astretsov19}. 

\section{Conclusions}\label{sec:conclusions}
We have presented a method for the fully self-consistent calculation of vertex corrections at the one- and two-particle level.
Similar to the parquet equations, a crossing-symmetric subset of the vertex diagrams is constructed iteratively, starting from a fully irreducible four-point vertex.
The mutual screening and renormalization of charge, spin, and particle-particle fluctuations is taken into account.
Fermionic/bosonic propagators and the vertex function are renormalized on an equal footing,
leading to an unbiased treatment of competing fluctuations from different channels. 

However, in contrast to the parquet equations, the vertex diagrams are classified according to the single-boson exchange (SBE) decomposition~\cite{Krien19-2},
which allows us to cast the self-consistent set of equations into three coupled integral equations for the Hedin three-leg vertex
and four simple update formulas for the self-energies of the fermionic and bosonic propagators.
This avoids the memory-intensive storage of intermediate four-point vertices and the matrix inversion of the Bethe-Salpeter equations. As a consequence, the resulting self-consistent equations in the SBE formalism are free from the vertex divergences of the two-particle self-energy~\cite{Schaefer13,Thunstroem18,Chalupa18}, which instead represent an additional complication for the numerical solution of the parquet equations. 

The parquet-like SBE equations for the Hedin vertex represent 
a \textit{practical} tool to get the enormous complexity of the two-particle correlations under control.
In fact, the presented applications for two zero-dimensional systems, 
the atomic limit and the Anderson impurity model,
can be converged within minutes on a single processing unit, whereas even for these relatively simple
problems the solution of the traditional parquet equations requires parallel programming and supercomputing capacity. 
The generalization of the SBE equations to quantum lattice models is appealing, because a finite-size scaling analysis is feasible. 
We defined a suitable approximation inspired by the D$\Gamma$A~\cite{Toschi07} which will allow investigations of the lattice Hubbard model within this novel parquet-like formalism.

\acknowledgments
We thank J. Mravlje for his reading of the manuscript and
M. Capone, K. Held, F. Kugler, A.I. Lichtenstein, J. Otsuki, A. Toschi, and C. Weber for discussions. 


\appendix
\section{Parquet-like equation for the singlet Hedin vertex}\label{app:parquetlike}
We derive Eq.~\eqref{eq:lambdaschannel} in the main text.
To this end, we use the following relation between the particle-hole channels $\alpha=\ch, \sz$ and the singlet particle-particle channel,
\begin{align}
    \varphi^{\pp,\sing}_{\nu\nu'\omegap}=\frac{1}{2}\left(\varphi^{\pp,\ch}_{\nu\nu'\omegap}-3\varphi^{\pp,\sz}_{\nu\nu'\omegap}\right).
\end{align}
We multiply by $\frac{1}{2}g_{\nu}g_{\omegap-\nu'}$, sum over $\nu'$ and add $-1$ on both sides, leading to,
\begin{align}
    \bar{\lambda}^{\sing}_{\nu\omegap}=-1+\frac{1}{4}\sum_{\nu'}\left(\varphi^{\pp,\ch}_{\nu\nu'\omegap}-3\varphi^{\pp,\sz}_{\nu\nu'\omegap}\right)g_{\nu}g_{\omegap-\nu'},
\end{align}
where we used definition~\eqref{eq:lambdafromphirightpp} to identify the singlet Hedin vertex $\bar{\lambda}^{\sing}$.
Now we insert the SBE decomposition~\eqref{eq:schannel} on the right-hand-side and are left with,
\begin{align}
    \bar{\lambda}^{\sing}_{\nu\omegap}=&\bar{\lambda}^{\firr,\sing}_{\nu\omegap}\label{app:parquetsing}\\
    +&\frac{1}{4}\sum_{\nu'}\Big(\nabla^{\ph,\ch}_{\nu\nu',\omegap-\nu-\nu'}+\nabla^{\phv,\ch}_{\nu\nu',\omegap-\nu-\nu'}-2U^\ch\notag\\
    -&3[\nabla^{\ph,\sz}_{\nu\nu',\omegap-\nu-\nu'}+\nabla^{\phv,\sz}_{\nu\nu',\omegap-\nu-\nu'}-2U^\sz]\Big)g_{\nu}g_{\omegap-\nu'},\notag
\end{align}
where we used definition~\eqref{eq:lambdafirrsing} of the fully $U$-irreducible singlet vertex $\bar{\lambda}^{\firr,\sing}$.
Using the definitions~\eqref{eq:nablaph} and~\eqref{eq:nablav} of the $U$-reducible vertices $\nabla^{\ph}$ and $\nabla^{\phv}$
one sees that they 
contribute equally to Eq.~\eqref{app:parquetsing}, leading to Eq.~\eqref{eq:lambdaschannel}.

\section{Decoupling ambiguity}\label{app:fierz}
We show that the (Fierz) decoupling ambiguity affects only the equation of motion, not the leading term of the two-particle self-energy.
The Hubbard interaction operator can be rewritten as ($0\leq r\leq1$),
\begin{align}
    Un_\uparrow n_\downarrow= U\frac{rnn+(r-1)mm}{2}-\left(r-\frac{1}{2}\right)Un,\label{eq:fierz}
\end{align}
where $n=n_\up+n_\dn, m=n_\up-n_\dn$ and we consider without loss of generality the `Ising' decoupling.
Using Eq.~\eqref{eq:fierz} one derives via the equation of motion $\partial_\tau g(\tau)$
a relation between the single-particle self-energy $\Sigma$, the screened interaction $w$, and the Hedin vertices $\lambda$,
as defined in Eqs.~\eqref{eq:w} and~\eqref{eq:hedinvertex},
\begin{align}
    \Sigma_\nu=&\left(\frac{1}{2}-r\right)U+rU\langle n\rangle\label{eq:hedinfierz}\\
    -&\frac{1}{\beta}\sum_\omega\left[rg_{\nu+\omega}w^\ch_\omega\lambda^{\ch}_{\nu\omega}+(1-r)g_{\nu+\omega}w^\sz_\omega\lambda^{\sz}_{\nu\omega}\right].\notag
\end{align}
The exact solution of the AIM~\eqref{eq:aim} satisfies this equation for arbitrary $r$, but approximations in general depend on this parameter.
However, for $r\neq\frac{1}{2}$ the decoupling leads to a shift of the chemical potential, as is evident from Eq.~\eqref{eq:fierz}.
In particular, Eq.~\eqref{eq:hedinfierz} requires a non-trivial cancellation between the frequency-independent terms
$(\frac{1}{2}-r)U+rU\langle n\rangle$ and the vertex corrections given by $\lambda$ to recover the Hartree energy $\frac{1}{2}U\langle n\rangle$. 
The requirement of a cancellation between different approximation levels is in general undesirable,
the decoupling ratio $r$ should therefore be set to $\frac{1}{2}$, see also Ref.~\cite{Zamani16}.

The bare interaction, i.e., the leading term of the two-particle self-energy, is not affected by the decoupling ambiguity.
This can be seen by an explicit derivation of the two-particle self-energy at RPA level, 
\begin{equation}
 \gamma^{0,\sigma'\sigma}=\frac{\delta\Sigma^H_\sigma}{\delta g^H_{\sigma'}},
\end{equation}
where $\Sigma^H_\sigma$ and $g^H_\sigma$ correspond to the Hartree approximation for a model with the interaction on the right-hand-side of Eq.~\eqref{eq:fierz}.
After a trivial calculation one readily confirms that the result
$\gamma^{0,\ch/\sz}=\gamma^{0,\up\up}\pm\gamma^{0,\up\dn}=\pm U$ is \textit{independent} of the decoupling ratio $r$~\cite{Krienthesis}. The leading term of the exact two-particle self-energy also has this property. 
Therefore, the definition of the irreducible three-leg vertex in equation~\eqref{eq:hedinvertex} is unique (independent of $r$),
which was used in Refs.~\cite{Krien19} and~\cite{Krien19-2} to derive the SBE decomposition.

\section{Exact Hedin vertex in the atomic limit}\label{app:hedin} 
The reducible three-leg particle-hole vertex can be obtained 
through the Ward identities~\cite{Krien18,vanLoon18}. 
At $\omega=0$ one has 
\begin{align} 
 g_{\nu}g_{\nu} \lambda^{\ch,\red}_{\nu,\omega=0} &= -\frac{d g_{\nu}}{d\mu}, \label{eq:wardch}\\
 g_{\nu}g_{\nu} \lambda^{\sz,\red}_{\nu,\omega=0} &= -\frac{d g_{\nu}}{d h},  \label{eq:wardsz}
\end{align}
where $\mu$ is the chemical potential and $h$ is an external magnetic field along the spin quantization axis, which couple linearly with the density $n=n_\up+n_\dn$ and the magnetization $m=n_\up-n_\dn$, respectively. At finite frequency, one instead has 
\begin{equation}
 g_{\nu} g_{\nu+\omega} \lambda^{\alpha,\red}_{\nu\omega} = -\frac{g_{\nu+\omega}-g_{\nu}}{\imath\omega}, \label{eq:wardw}
\end{equation}
and the irreducible (Hedin) vertex is hence given by~\cite{Krien19} 
\begin{equation}
 \lambda^{\alpha}_{\nu\omega} = \frac{\lambda^{\alpha,\red}_{\nu\omega}}{1+\frac{1}{2}\chi^{\alpha}_{\omega}U^{\alpha}},
\end{equation}
with $\alpha=\ch,\sz$. At half-filling, the particle-particle vertex in the singlet channel is obtained by symmetry from the particle-hole one as $\lambda^{\sing}_{\nu\omega}=-\lambda^{\ch}_{\nu,-\omega}$~\cite{Krien19-2}.

In the AL, since the exact form of Green's function is known, it is possible to evaluate the right-hand-side of Eqs.~\eqref{eq:wardch},~\eqref{eq:wardsz},~and~\eqref{eq:wardw} analytically. 
The full expressions can be found, e.g., in Ref.~\cite{Krien19}. 

\section{Symmetries}\label{app:symmetry}
We prove that the SBE decomposition~\eqref{eq:jib_full} satisfies the exact crossing-symmetry of the vertex function,
provided that this symmetry holds for the fully irreducible vertex $\varphi^\firr$ together with a symmetry of a Hedin vertex.
\subsection{Symmetry of the Hedin vertices}
The Hedin vertices obey the symmetries,
\begin{align}
    \lambda^{\ch,\sz}_{\nu-\omega/2,\omega}=&\left(\lambda^{\ch,\sz}_{-\nu-\omega/2,\omega}\right)^*\label{eq:sym_ph},\\
    \lambda^{\sing}_{\nu+\omegap/2,\omegap}=&\lambda^{\sing}_{-\nu+\omegap/2,\omegap}.\label{eq:sym_pp}
\end{align}
The first relation follows from time-reversal symmetry~\cite{vanLoon14,Krien19-2}.
It implies that in the particle-hole channels the Hedin vertex is symmetric around the point $-\omega/2$ as a function of $\nu$,
cf. Figs.~\ref{fig:hedin_AL} and~\ref{fig:hedin_AIM}.
The second relation is a consequence of the crossing-symmetry of the \textit{exact} singlet vertex function~\cite{Rohringer12},
$f^\sing_{\nu\nu'\omegap}=f^\sing_{\nu,\omegap-\nu',\omegap}$, and hence this vertex is symmetric around $\omegap/2$.
\subsection{Crossing-symmetry}
The exact vertex satisfies the crossing-symmetry, which reads for the particle-hole channels~\cite{Rohringer12,Krien19-2},
\begin{align}
    f^{\alpha}_{\nu\nu'\omega}
    \!=&-\!\frac{1}{2}\!\left(f^{\ch}_{\nu,\nu+\omega,\nu'-\nu}\!+\![3\!-\!4\delta_{\alpha,\sz}]f^{\sz}_{\nu,\nu+\omega,\nu'-\nu}\right).\label{eq:crossing}
\end{align}
However, when we solve the self-consistent cycle in Fig.~\ref{fig:cycle}, at an intermediate step the crossing-symmetry could be violated.
In the evaluation of the traditional parquet approximation, enforcing the crossing-symmetry at every step is crucial for the numerical stability~\cite{Tam13}.
Furthermore, one should make sure that this symmetry holds even when we use an approximation for the fully irreducible vertex $\varphi^\firr$.

Indeed, one sees that Eq.~\eqref{eq:crossing} holds by inserting the SBE decomposition~\eqref{eq:jib_full} on both sides,
provided that $\varphi^\firr$ satisfies Eq.~\eqref{eq:crossing} [in place of $f$],
and when Eq.~\eqref{eq:sym_pp} holds for the Hedin vertex $\lambda^\sing$ of the singlet channel.
In a self-consistent calculation of the Hedin vertices, we can therefore enforce the crossing-symmetry by imposing Eq.~\eqref{eq:sym_pp}.
Using also Eq.~\eqref{eq:sym_ph} halves the numerical cost.

\section{Exact relations}\label{app:exactrelations}
We discuss whether several properties of the exact solution are satisfied by approximations based on the three-leg parquet.
By virtue of the one- and two-particle self-consistency, the potential energy, given by the Migdal-Galitskii formula~\cite{Galitskii58},
$E_\text{pot}=\sum_\nu g_\nu\Sigma_\nu$, is consistent with the two-particle level.
To see this, we multiply Eq.~\eqref{eq:hedin} with $g_\nu$ and sum over $\nu$~\footnote{
The summation $\sum_\nu$ implies a convergence factor $e^{\imath\nu0^+}$, see also Ref.~\cite{Krien17}.},
\begin{align}
    E_\text{pot}=&\sum_\nu g_\nu\Sigma_\nu\notag\\
    =&\frac{U\langle n\rangle^2}{4}-\frac{1}{2}\sum_{\nu\omega} g_\nu g_{\nu+\omega}\left[w^\ch_\omega\lambda^\ch_{\nu\omega}+w^\sz_\omega\lambda^\sz_{\nu\omega}\right]\notag\\
    =&\frac{U\langle n\rangle^2}{4}-\frac{1}{4}\sum_\omega \left[U^\ch\chi^\ch_\omega+U^\sz\chi^\sz_\omega\right].\label{eq:epot}
\end{align}
In the last step we used Eq.~\eqref{eq:pi_ph} for the polarization $\pi$ and $w^\alpha_\omega\pi^\alpha_{\omega}=\frac{1}{2}U^\alpha \chi^\alpha_\omega$,
which follows from Eqs.~\eqref{eq:w} and~\eqref{eq:wph}. Using $U^\ch=-U^\sz=U$, Eq.~\eqref{eq:epot} implies that the approximation satisfies an exact
relation between the susceptibility $\chi$ and $E_\text{pot}$~\cite{Krien17}, and that the latter is consistent with the Migdal-Galitskii formula.

Next, we consider the 
{asymptotic behavior} of the self-energy. To this end, we expand Eq.~\eqref{eq:hedin}
retaining terms up to order $\mathcal{O}(\frac{1}{\nu})$, {to obtain} 
\begin{align}
    \Sigma_\nu=&\frac{U\langle n\rangle}{2}-\frac{U^2}{4\imath\nu}\sum_\omega \left[\chi^\ch_\omega+\chi^\sz_\omega\right]+\mathcal{O}\left(\frac{1}{\nu^2}\right).\label{eq:asymptote}
\end{align}
Here we used that $\lambda_{\nu\omega}\rightarrow1$ for large $\nu$~\footnote{We need not consider the case $\nu\approx{-\omega}$ separately,
since it implies that both $\nu$ and $\omega$ are large and in turn $\lambda_{\nu\omega}\rightarrow 1$.} and Eq.~\eqref{eq:w}.
This relation expresses a further consistency 
of the one-particle level with $\chi$.
However, Eq.~\eqref{eq:asymptote}, does not automatically guarantee that the asymptotic coefficient, i.e., the first moment of the self-energy, assumes the exact value~\cite{Rohringer16},
{$\Sigma_\nu^{(1)} = \frac{1}{2\imath\nu} \langle n\rangle U^2 \left[1-\frac{\langle n\rangle}{2}\right]$}.

Like other parquet schemes, the presented approach implies the crossing-symmetry of the vertex (see Appendix~\ref{app:symmetry}), but the Ward identity can be violated. 
{On the other hand, conserving approximations in the sense of Baym and Kadanoff do not usually satisfy the crossing-symmetry. 
Hence, conserving and crossing-symmetric approaches can be considered to be complementary}~\cite{Bickers04,Krienthesis}.

\bibliographystyle{apsrev}
%

\end{document}